\documentclass[11pt]{article}      
\usepackage{graphicx}
\usepackage{enumerate}
\usepackage[small,bf,hang]{caption} 
\usepackage{subcaption}
\usepackage{amsmath}
\usepackage{amssymb}
\usepackage{rotating}
\usepackage{authblk}
\usepackage{xcolor}		 	
\usepackage{verbatim}		
\usepackage{colortbl}
\usepackage{booktabs}
\usepackage{dcolumn}
\usepackage{expdlist}  
\usepackage{float}
\usepackage{url}
\usepackage{fullpage}

\usepackage{longtable}
\usepackage{threeparttable}
\usepackage{threeparttablex}
\usepackage[ruled,vlined]{algorithm2e}

\usepackage{mathptmx}   
   
\newtheorem{lemma}{Lemma}
\newtheorem{definition}{Definition}
\newtheorem{theorem}{Theorem}

\SetKwFunction{Range}{range}
\newcommand{\forcond}{$times$ \KwTo $\Range{maxAmountInitialSpanningTrees}$}
\SetKwProg{FnA}{Functie}{}{end}\SetKwFunction{FRecursA}{makeTree}%
\SetKwProg{FnB}{Functie}{}{end}\SetKwFunction{FRecursB}{obtainInitialSpanningTree}%
\SetKwProg{FnC}{Functie}{}{end}\SetKwFunction{FRecursC}{estimateHCN}%
\SetAlgoLongEnd

\SetKwFunction{Range}{range}
\SetKwFor{While}{while}{:}{fintq}%
\SetKw{KwTo}{in}\SetKwFor{For}{for}{\string:}{}%
\newcommand{\forcondA}{$path$  \KwTo{$newPathPartition$}}
\newcommand{\forcondB}{$edge$  \KwTo{$minWeightHamiltonianCycle$}}

\begin{document}
	
	\title{A multi-start local search algorithm for the Hamiltonian completion problem on undirected graphs}
	\author{Jorik Jooken, Pieter Leyman, Patrick De Causmaecker} 
	\affil{\small KU Leuven Kulak, Department of Computer Science, CODeS, Etienne Sabbelaan 53, 8500 Kortrijk (Belgium)\\ jorik.jooken@kuleuven.be, pieter.leyman@kuleuven.be, patrick.decausmaecker@kuleuven.be}
	\date{}
	\maketitle
	
	\begin{abstract}
		This paper proposes a local search algorithm for a specific combinatorial optimisation problem in graph theory: the Hamiltonian Completion Problem (HCP) on undirected graphs. In this problem, the objective is to add as few edges as possible to a given undirected graph in order to obtain a Hamiltonian graph. This problem has mainly been studied in the context of various specific kinds of undirected graphs (e.g. trees, unicyclic graphs and series-parallel graphs). The proposed algorithm, however, concentrates on solving HCP for general undirected graphs. It can be considered to belong to the category of matheuristics, because it integrates an exact linear time solution for trees into a local search algorithm for general graphs. This integration makes use of the close relation between HCP and the minimum path partition problem, which makes the algorithm equally useful for solving the latter problem. Furthermore, a benchmark set of problem instances is constructed for demonstrating the quality of the proposed algorithm. A comparison with state-of-the-art solvers indicates that the proposed algorithm is able to achieve high-quality results.
	\end{abstract}

	
	\section{Introduction}
	\label{intro}
	The problem of determining whether a given undirected graph contains a Hamiltonian cycle is one of Karp's famous 21 NP-complete problems \cite{Karp:1972}. Being such a famous problem, it has been studied intensively for many years by the research community. This research has also led to many variations of this problem. One such variation is an interesting generalisation of the original problem in an optimisation context: the Hamiltonian Completion Problem (HCP), which is the main topic of this paper.
	
	In this paper, we develop a matheuristic \cite{Boschetti:2009} for HCP on undirected graphs. We will identify a restricted version of this problem that can be solved exactly in polynomial time. The solution of this restricted problem is integrated into a multi-start local search algorithm that attempts to solve HCP on general undirected graphs. This algorithm will repeatedly perturb the solution of the restricted problem to obtain a high-quality solution for the original problem.
	
	The remainder of this paper is organised as follows: in section \ref{problemDescription} a formal description of HCP is given. We further motivate this problem by discussing applications in section \ref{applications}. Section \ref{literatureOverview} contains a brief overview of earlier work that has been done for this problem. We highlight those results from literature that are relevant to understand the remainder of this paper. In section \ref{localSearch} the multi-start local search algorithm that we propose for solving HCP is explained. The explanation is built in a bottom-up fashion: we first describe the separate components of the algorithm, after which the complete algorithm is explained. In section \ref{benchmark} we will describe how a benchmark set of problem instances is generated. These problem instances will be used to compare the algorithm from this paper with state-of-the-art algorithms. The results of these experiments are described in section \ref{results}. Finally, a conclusion and possible avenues for further work are discussed in section \ref{conclusionsFurtherWork}.
	
	\section{Problem description}
	\label{problemDescription}
	
	The Hamiltonian Completion Problem (HCP) on undirected graphs \cite{Goodman:1974} is the following problem: given an undirected graph $G=(V,E)$, find a set of edges $E'$ such that $G'=(V,E \cup E')$ is a Hamiltonian graph and the size of $E'$ is as small as possible. A Hamiltonian graph is a graph that has a Hamiltonian cycle (a cycle that visits each node exactly once, except for the first and the last node, which are visited twice).
	
\begin{figure*}[h!]
		\centering
		\includegraphics[width=0.9\textwidth]{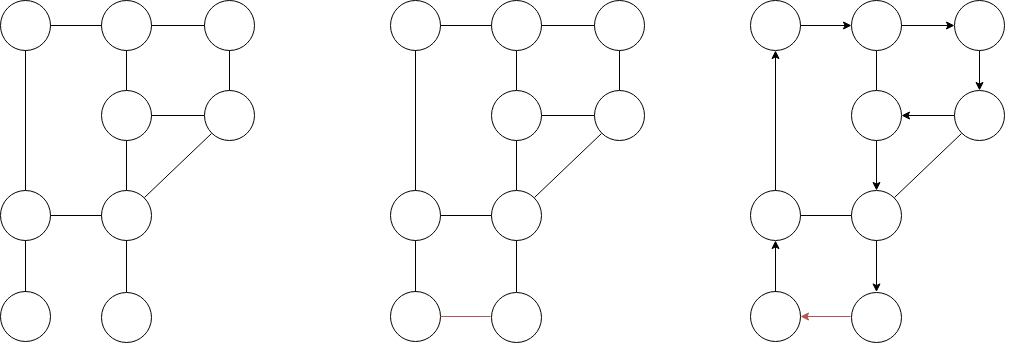}
		\caption{Graph $G$, graph $G'$ and a Hamiltonian cycle in $G'$}
		\label{HamiltonianCompletionProblem}       
	\end{figure*}

	As an example, consider Fig. \ref{HamiltonianCompletionProblem}. In this figure, the graph on the left is the graph $G$ for which HCP has to be solved. The graph in the middle shows $G'$, which is obtained by adding $1$ edge to $G$. The graph on the right illustrates that $G'$ is indeed a Hamiltonian graph (by showing the existence of a Hamiltonian cycle).
	
	\section{Applications}
	\label{applications}
	The need for solving HCP arises naturally in many applications. A first example application involving frequency assignment to transmitters has been identified in \cite{Franzblau:2002}. HCP can also be used to solve a special case of the Travelling Salesman Problem (TSP) \cite{Applegate:2007}. A problem instance of TSP on a complete edge-weighted graph with only two different edge weights can be solved by solving HCP, as discussed in \cite{Rayward:1987}.
	
	Yet another application comes from a generalisation of the Bottleneck Travelling Salesman Problem \cite{Gilmore:1964}. In this generalisation, we are given an integer $k$ and an edge-weighted undirected graph $G=(V,E)$. The goal is to find a set of at most $k$ paths in $G$ such that every node in $V$ belongs to exactly one path and such that the maximum edge weight amongst all edges in the paths is minimised. We assume that $k$ is chosen such that at least one solution exists. This problem can be solved by doing a binary search for the smallest maximum edge weight. To determine for a specific weight $w$ whether a solution exists such that the maximum edge weight is at most $w$, one has to solve an instance of HCP. More specifically, let $f(G,w)$ denote an unweighted, undirected graph obtained by removing from $G$ those edges that have a larger weight than $w$. The weights of the remaining edges are ignored to make the graph unweighted. The set of nodes in $f(G,w)$ is the same as the set of nodes in $G$. Let $E'$ be the set of edges that is obtained by solving HCP on $f(G,w)$ (i.e. adding the edges in $E'$ to $f(G,w)$ results in a Hamiltonian graph). Now, there exists a solution with maximum edge weight at most $w$ if and only if $E'$ contains at most $k$ edges. Let $w^*$ be the smallest value for which $E'$ contains at most $k$ edges and let $H$ be the graph obtained by adding the edges in $E'$ to $f(G,w^*)$ (so H is Hamiltonian). Now the set of at most $k$ paths in $G$ (the answer to the problem) can be found by taking any Hamilonian cycle in $H$ and dropping those edges that are in $E'$. The required set of paths is simply the obtained set of connected components. All edges in the obtained paths will have a weight that is at most equal to $w^*$.
	
	\section{Literature overview}
	\label{literatureOverview}
	
	It is clear that HCP on undirected graphs is NP-hard, because it can be used to determine whether a graph has a Hamiltonian cycle. Determining whether a graph has a Hamiltonian cycle, known as the Hamiltonian cycle problem, is NP-complete \cite{Garey:1990}. This means that HCP on undirected graphs cannot be solved exactly in polynomial time, unless $P=NP$. Hence, most of the existing literature concentrates on efficiently solving HCP for special cases of undirected graphs. Amongst those special cases are algorithms that solve HCP in polynomial time for trees and unicyclic graphs \cite{Goodman:1974}\cite{Goodman:1975}\cite{Franzblau:2002}, line graphs of trees \cite{Agnetis:2001}\cite{Raychaudhuri:1995}, line graphs of cacti \cite{Detti:2004} and series-parallel graphs \cite{Takamizawa:1982}. HCP has also been studied in the context of sparse random graphs \cite{Gamarnik:2005}. The amount of literature that concentrates on heuristically solving HCP is more limited \cite{Rayward:1987}\cite{Detti:2007}\cite{Maretic:2015}.
	
	In the remainder of this section, we will introduce some definitions and results from existing literature that will be needed to understand the rest of the paper.

	\subsection{Reduction to Travelling Salesman Problem}
	\label{reductionTSP}
	Lemma \ref{TSPlemma} \cite{Detti:2007} states that HCP on undirected graphs can be solved by solving an instance of the symmetric travelling salesman problem. Let $G=(V,E)$ be an undirected, unweighted graph. Let $G'=(V',E')$ be an undirected, weighted, complete graph such that $V=V'$ and $E'$ consists of $\frac{|V|*(|V|-1)}{2}$ edges linking all pairs of nodes in $V'$. The weight of an edge $e \in E'$ is defined to be equal to $0$ if and only if there is also an edge in G that links the same pair of nodes and otherwise the weight is equal to $1$.
	
	\begin{lemma}
		\label{TSPlemma}
		HCP on $G$ can be solved by determining the minimum weight Hamiltonian cycle in $G'$ and adding to $G$ those edges of this cycle that have weight $1$. Determining the minimum weight Hamiltonian cycle in $G'$ is an instance of the symmetric travelling salesman problem.
	\end{lemma}
	
	Using this lemma, HCP can be solved by reducing it to an instance of the symmetric TSP problem and then using state-of-the-art TSP solvers like Concorde \cite{Applegate:2001} (an exact solver) or LKH \cite{Helsgaun:2000} (a heuristic solver). When the graphs for which we want to solve HCP become bigger however, the computation time needed to produce high-quality solutions for TSP with these solvers might no longer be acceptable.
	
	\subsection{Minimum path partition}
	\label{minimumPathPartition}
	In this subsection the minimum path partition problem is introduced. This problem is closely related to HCP and will be an important concept needed to understand the algorithm in section \ref{localSearch}. First, some terminology is introduced.
	
	\begin{definition}
		\label{vertexDisjoint}
		Two paths $P_1$ and $P_2$ are called vertex-disjoint if they do not have a vertex in common. 
	\end{definition}
	
	\begin{definition}
		\label{pathPartition}
		A path partition of a graph $G=(V,E)$ is a set of pairwise vertex-disjoint paths $\{P_1,P_2,\ldots,P_k\}$ such that every node in $V$ belongs to exactly $1$ path $P_i$.
	\end{definition}
	
	\begin{definition}
		\label{minimumPathPartition}
		A minimum path partition of graph $G=(V,E)$ is a path partition of $G$, which consists of the least amount of paths over all path partitions of $G$.
	\end{definition}
	
	\begin{definition}
		\label{pathPartitionNumber}
		The path partition number of a graph G (denoted $PPN(G)$) is defined as the amount of paths in a minimum path partition of G. 
	\end{definition}
	
	\begin{definition}
		\label{HamiltonianCompletionNumber}
		The Hamiltonian completion number of a graph $G$ (denoted $HCN(G)$) is defined as the minimum amount of edges that need to be added to $G$ to obtain a Hamiltonian graph.
	\end{definition}
	
	The minimum path partition problem on undirected graphs asks to find a minimum path partition on a given undirected graph. The following lemma \cite{Goodman:1974} shows that this problem is closely related to HCP on undirected graphs:
	
	\begin{lemma}
		\label{HCNPPNLemma}
		Let $G=(V,E)$ be an undirected graph. If $G$ is Hamiltonian, then $HCN(G)=0$ and $PPN(G)=1$. Else, $HCN(G)=PPN(G)$.
	\end{lemma}
	
\begin{figure*}[h!]
		\centering
		\includegraphics[width=0.9\textwidth]{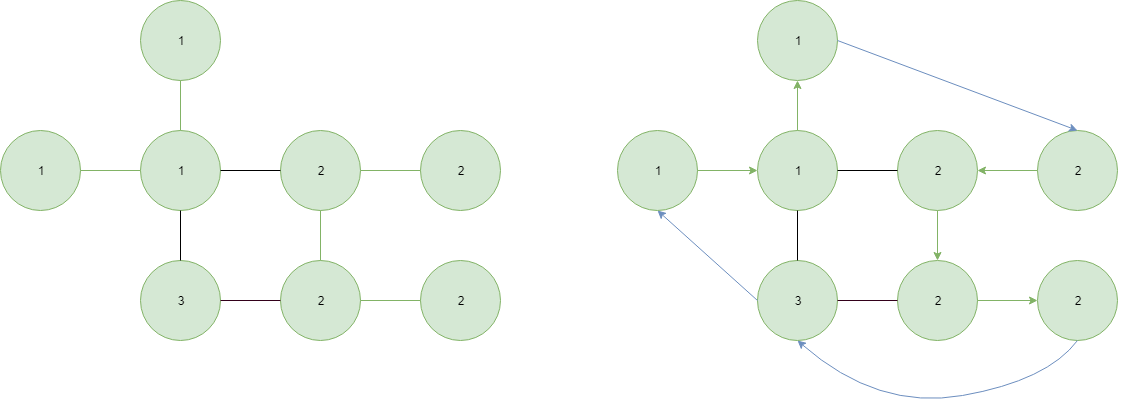}
		\caption{Graph $G$ and a Hamiltonian cycle in graph $G'$}
		\label{PathPartitionToHamCyc}       
	\end{figure*}

	In case $PPN(G)>1$, a minimum path partition of $G$ can be used to determine which edges to add to $G$ to make it Hamiltonian (by cyclically linking the endpoints of the paths of the path partition). This is illustrated in Fig. \ref{PathPartitionToHamCyc}. The graph on the left of the figure is the graph $G$ for which HCP needs to be solved. The minimum path partition is indicated on this graph. Nodes with an equal number are part of the same path. The edges in green indicate the links between nodes that belong to the same path. Hence, the path partition number of the shown graph is equal to 3. The graph on the right of the figure illustrates that a Hamiltonian graph can be obtained by cyclically linking the endpoints of the paths of the path partition. A Hamiltonian cycle in this graph is indicated.
	
	In case $PPN(G)=1$, either $HCN(G)=0$ or $HCN(G)=1$, but there is no simple rule to deduct from a given minimum path partition which of these two equalities is the correct one. Note that $HCN(G)=0$ implies that $PPN(G)=1$ and that there exists at least one minimum path partition of $G$ such that there is an edge between the endpoints of the single path in that minimum path partition. However, not necessarily all minimum path partitions of $G$ have this property. Hence, one cannot use the (non-)existence of an edge between the endpoints of a path to determine whether $HCN(G)=0$ or $HCN(G)=1$.
	
	An immediate consequence of lemma \ref{HCNPPNLemma} and the corresponding construction is the following lemma:
	
	\begin{lemma}
		\label{separationLemma}
		Let $G=(V,E)$ be an undirected graph that consists of multiple components (i.e. $G$ is a disconnected graph). HCP for $G$ can be solved by computing a minimum path partition for every component of $G$ separately.
	\end{lemma}
	
	Using this lemma, it is clear that HCP for a graph $G$ can be solved by solving HCP for every component of $G$ separately. Hence, in the rest of this paper we will assume that the graph $G$ for which HCP has to be solved, is a connected graph, unless stated otherwise.
	
	\subsection{Special properties for trees}
	\label{trees}
	
	If we restrict the input graph $G$, for which HCP has to be solved, to be a tree, we can efficiently solve HCP \cite{Goodman:1975}\cite{Franzblau:2002}:
	
	\begin{lemma}
		\label{treeLemma}
		Let $G=(V,E)$ be a tree. It is possible to compute a minimum path partition of $G$ with a time complexity of $O(|V|)$.
	\end{lemma}
	
	The minimum path partition can then be used to solve HCP, as seen in the previous subsection. The algorithm to compute the minimum path partition of a tree is not further discussed in this paper, but interested readers are referred to \cite{Goodman:1975} and \cite{Franzblau:2002}. 
	
	Furthermore, there is a relation between the PPN of a graph and the PPN of its spanning trees \cite{Goodman:1974}:
	
	\begin{lemma}
		\label{spanningTreeLemma}
		Let $G=(V,E)$ be an undirected graph. For all spanning trees $T$ of $G$, the following inequality holds: $PPN(T) \geq PPN(G)$. Furthermore there is at least one spanning tree $T$ of $G$ such that $PPN(T)=PPN(G)$.
	\end{lemma}
	
	By combining lemma \ref{treeLemma} and \ref{spanningTreeLemma}, it is clear that finding a minimum path partition for a graph can be done by finding a minimum path partition for all of its spanning trees. Unfortunately, this approach is not always feasible, because the amount of spanning trees of a graph can be very large (e.g. a complete graph with $n$ vertices has $n^{n-2}$ spanning trees \cite{Cayley:2009}). However, note that the fact that a graph has many spanning trees does not necessarily imply that it is hard to find a minimum path partition for this graph. As can be seen in the results section, the algorithm that is proposed in this paper often is able to find the minimum path partition for graphs that have many spanning trees.
	
	\section{Multi-start local search algorithm}
	\label{localSearch}
	
	The algorithm that we propose to solve HCP on an undirected graph $G$ will attempt to find a minimum path partition of $G$. It does this indirectly, however, by trying to search for an optimal spanning tree $T$ of $G$ (i.e. such that $PPN(T)=PPN(G)$). The algorithm starts with multiple initial spanning trees, that will be perturbed in order to obtain better spanning trees. A spanning tree $T_1$ is called better than a spanning tree $T_2$ if $PPN(T_1)$ is smaller than or equal to $PPN(T_2)$. 
	
	\subsection{Initial spanning tree}
	\label{initialSpanningTree}
	
	Several steps are taken to obtain an initial spanning tree of the graph $G=(V,E)$:
	
	\begin{enumerate}[Step 1:]
		\item HCP is first transformed into a symmetric TSP instance, as discussed in section \ref{reductionTSP}. A heuristic solution for this symmetric TSP instance is then computed by the heuristic TSP solver LKH in order to obtain an approximation for the minimum weight Hamiltonian cycle in the complete graph $G'$ (where $G'$ is the graph as described before lemma \ref{TSPlemma}). In order to keep the running time low, the parameters that are supplied to LKH in this step are purposely chosen in such a way that only a fraction of the full power of the LKH solver is being used (e.g. the supplied parameter \textit{RUNS} is much smaller than the default amount of runs, see section \ref{tuning}).
		\item This Hamiltonian cycle in $G'$ is used to obtain a path partition of $G$ (by omitting those edges of the cycle that have weight $1$). 
		\item This path partition is modified to obtain another path partition, by applying a rotation move \cite{Posa:1976} to all of its paths (in case it is possible to apply such a move). A rotation move modifies a path in the following way: let $P=(v_1,v_2,\ldots,v_k)$ be a path consisting of $k$ vertices. If there is a vertex $v_i$ (with $1<i<k-1$) such that there is an edge between $v_i$ and $v_k$ in $G$, a rotation move can be applied to obtain the path $P'=(v_1,v_2,\ldots,v_i,v_k,v_{k-1},v_{k-2},\ldots,v_{i+1})$. If there are multiple choices for $v_i$, an arbitrary one is chosen. Fig. \ref{rotationMove} illustrates the situation before step 3 is executed (on the left) and afterwards (on the right). A rotation move is applied for the first path (indicated with number 1). 
		
		\begin{figure*}[h!]
			\centering
			\includegraphics[width=0.85\textwidth]{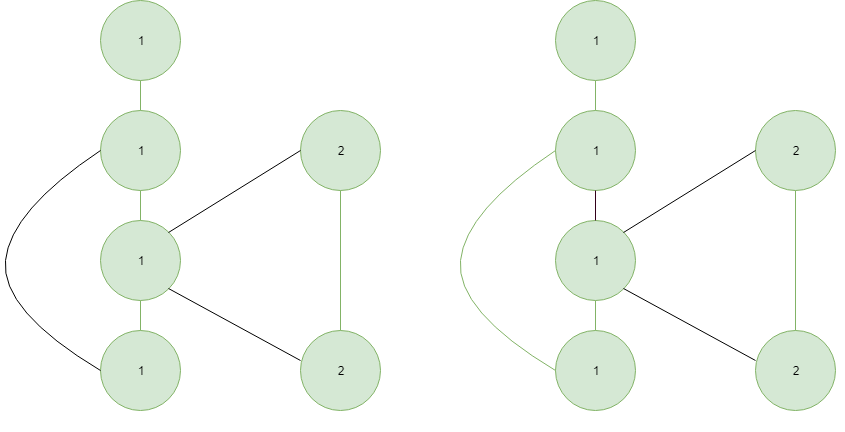}
			\caption{The paths before and after applying rotation moves. The edges in green indicate the links between neighboring nodes in a path.}
			\label{rotationMove}       
		\end{figure*}
		
		\item The obtained path partition is transformed into a spanning tree of $G$ by adding a set of edges of $G$, that do not already belong to the paths in the path partition. If the path partition consists of $k$ paths, $k-1$ edges have to be added in order to obtain a spanning tree. Not all edges are added with equal probability. The edges are grouped in two categories: preferred edges and normal edges. Preferred edges are edges that have at least one endpoint that is also an endpoint of some path in the given path partition. The other edges are called normal edges. The edges are added one by one. Every time an edge needs to be added, a decision is made whether to add an edge from the category of preferred edges or normal edges. If one of the categories does not contain any edges, a random edge from the other category is selected. Else, a random edge is chosen from the category of preferred edges with probability $\frac{preferredRatio}{preferredRatio+1}$ (and from the category of normal edges with probability $1-\frac{preferredRatio}{preferredRatio+1}=\frac{1}{preferredRatio+1}$ ). By defining the probabilities in this way, we have the property that the parameter \textit{preferredRatio} indicates the ratio between the probability that an edge is selected from the category of preferred edges and the probability that an edge is selected from the category of normal edges. This parameter can be used to indicate how many times more likely it is to select an edge from the category of preferred edges than an edge from of the category of normal edges. The spanning tree that is obtained, tends to have a lower path partition number when more edges from the category of preferred edges are added. Hence, the intended use of this parameter is to choose \textit{preferredRatio} such that it is larger than 1, but this is not explicitly necessary. 
	\end{enumerate}
	
	Algorithm 1 illustrates the pseudocode to obtain an initial spanning tree. Here, \textit{obtainInitialSpanningTree} is a function that accepts as input a graph $G$ and produces as output a spanning tree of $G$. The function \textit{solveTSP} is a function that accepts as input a graph and produces as output an approximation of the minimum weight Hamiltonian cycle in that graph. The function \textit{makeTree} executes step 3 and step 4 from above. It accepts as input a graph $G$ and a path partition of $G$. It produces as output a spanning tree of $G$. The pseudocode of this function is given in algorithm 2. The function \textit{applyRotationMove} accepts as input a path $p$ and returns as output the path that is obtained by applying a rotation move on $p$. Finally, the function $addEdges$ corresponds to step 4 from above. It accepts as input a graph $G$ and a path partition of $G$. It produces as output a spanning tree of $G$.
	
	\begin{algorithm}[h!]
		\FnB{\FRecursB{G}}{
			\KwData{\\G: an undirected graph}
			\KwResult{\\A spanning tree of G}
			\tcc{Start of code}
			{Let $G'=(V',E')$ be a complete graph such that $V=V'$ and the weight of an edge $e \in E'$ is defined to be equal to 0 if there is also an edge in $G$ that links the same pair of nodes and 1 otherwise}\;
			\tcc{step 1}
			{$minWeightHamiltonianCycle \gets solveTSP(G')$}\;
			\tcc{step 2}
			{$pathPartition \gets \emptyset$}\;
			\For{\forcondB}{
				\uIf{$weight(edge)=0$}{
					{$pathPartition \gets pathPartition \cup edge$}\;
				}
			}
			\tcc{step 3 and 4}
			{\Return $makeTree(G,pathPartition)$}
		}
		\caption{Obtain an initial spanning tree of the graph}
	\end{algorithm}
	
	\begin{algorithm}[h!]
		\FnA(\tcc*[h]{Step 3 and 4}){\FRecursA{G, pathPartition}}{
			\KwData{\\G: an undirected graph\\pathPartition: a path partition of G}
			\KwResult{\\A spanning tree of G}
			\tcc{Start of code}
			{$newPathPartition \gets pathPartition$}\;
			\For{\forcondA}{
				\uIf{rotation move can be applied on $path$}{
					{$path \gets applyRotationMove(path)$}\;
				}
			}
			{\Return $addEdges(G,newPathPartition)$}
		}
		\caption{Transform a path partition into a spanning tree}
	\end{algorithm}

	To analyse the time complexity of obtaining an initial spanning tree, we will analyse the time complexity of the different steps separately. The time complexity of step 1 heavily depends on the parameters that are supplied to LKH. The parameters for LKH can be chosen such that the preprocessing step of this solver runs in nearly linear time (by choosing the \textit{POPMUSIC} option for candidate set generation \cite{Helsgaun:2018}) and the amount of local search steps is constant. For a more thorough treatment of the time complexity of this solver, the reader is referred to \cite{Helsgaun:2000}. It is clear that step 2 and step 3 can each be implemented in $O(|V|+|E|)$. In order to efficiently execute step 4, one has to be able to quickly verify whether adding a certain edge would result in a cycle (if this is the case, the edge should not be added). This is a well-known problem that can be efficiently solved by using the union-find data structure \cite{Tarjan:1984}. By using this data structure, one can verify in $O(\alpha(|V|))$ time whether adding an edge would result in a cycle. Here, $\alpha(|V|)$ denotes the inverse of the Ackermann function \cite{Ackermann:1928}. Because of this, step 4 can be implemented in $O(|E|*\alpha(|V|))$. Hence, the total time complexity to obtain an initial spanning tree is $LKHTime(V,E)+O(|V|+|E|*\alpha(|V|))$, where $LKHTime(V,E)$ denotes the time that the LKH solver spends in step 1.
	
	\subsection{Perturbing the spanning tree}
	\label{perturbing}
	
	The spanning tree that is obtained, is repeatedly perturbed in order to obtain better spanning trees. A single perturbation consists of multiple steps, which are explained below. In the following steps, we will denote by $G$ the graph for which HCP needs to be solved and by $T$ a spanning tree of $G$ that will be perturbed. The process of perturbing a spanning tree can be thought of as marking some edges as active and other edges as inactive. Initially, all edges in $G$ are inactive, except for the edges in $T$. After performing the perturbation, the set of active edges forms a (new) spanning tree of $G$. The steps of a single perturbation are explained below.
	
	\begin{enumerate}[Step 1:] 
		\item Compute a minimum path partition of $T$. Mark all the edges that are part of $T$, but not part of the path partition as inactive. Note that any path partition of $T$ is also a path partition of $G$.
		\item The main idea of this step is that we try to alter the path partition from the previous step to obtain a new path partition of $G$ that consists of fewer paths. One possible way to approach this problem would be to try to link together endpoints of the paths from the path partition by marking the connecting edges as active. However, this idea can be generalised, without resulting in a higher time complexity: let $PP =\{P_1,P_2,\ldots,P_k\}$ be the path partition obtained from step 1, which consists of $k$ paths. For all of the paths in $PP$, it is verified whether the path can be transformed into a cycle by linking together the two endpoints of the path (i.e. it is verified whether there exists an edge in $G$ such that the endpoints of this edge are equal to the endpoints of the path). In case the path can indeed be transformed into a cycle, this transformation is performed (the edge that is needed for this transformation is marked as active). After doing this, we have a set of paths $S_1$ (of size $k_1$) and a set of cycles $S_2$ (of size $k_2$) such that $k_1+k_2=k$. 
		
		These two sets will be iteratively updated by joining either a path with another path, a cycle with a path or a cycle with another cycle. To do this, the algorithm goes over all edges in $G$ and verifies whether the current edge can be used to join the structures (paths or cycles) at both endpoints of the edge. In case this is possible, the structures are joined and the sets $S_1$ and $S_2$ are updated. The three cases in which an edge can be used to join two structures are discussed next.
		
		\begin{enumerate}[C{a}se 1:]
			\item The edge is between two paths $P_i$ and $P_j$. The edge can be used to join $P_i$ and $P_j$ if and only if the edge is between an endpoint of $P_i$ and an endpoint of $P_j$ (and the paths $P_i$ and $P_j$ are not the same paths). If $P_i$ and $P_j$ can indeed be joined, the edge is marked as active, resulting in a longer path $P'$. The set $S_1$ is updated by removing $P_i$ and $P_j$ and adding $P'$.
			
			\item The edge is between a path $P_i$ and a cycle $C_j$. The edge can be used to join $P_i$ and $C_j$ if and only if exactly one endpoint of the edge corresponds to an endpoint of $P_i$. If $P_i$ and $C_j$ can indeed be joined, the edge is marked as active. Furthermore, an edge of $C_j$ is marked as inactive such that the joined structure forms a path. There are two possible such edges that could be marked as inactive. An arbitrary one of these two edges is chosen. By joining $P_i$ and $C_j$ a new path $P'$ is obtained. The set $S_1$ is updated by removing $P_i$ and adding $P'$. The set $S_2$ is updated by removing $C_j$.
			
			Fig. \ref{pathCycleJoin} illustrates case 2. The graph at the top of the figure consists of the path $1-2-3$ and the cycle $4-5-6-7-4$. The green and blue edges represent the links between consecutive nodes of the path and the cycle respectively. The black edge represents an inactive edge. There is an edge between the cycle and an endpoint of the path: the edge that links node 3 and node 4. This edge is used to join the path and the cycle and will be marked as active. There are two choices for an edge that could be marked as inactive in order to obtain a new path, namely the edge between node 4 and node 5 and the edge between node 4 and node 7. In this figure, the edge between node 4 and node 5 is marked as inactive in order to obtain the path $1-2-3-4-7-6-5$. The resulting path is shown at the bottom of the figure. 
			
			\begin{figure*}[h!]
				\centering
				\includegraphics[width=0.75\textwidth]{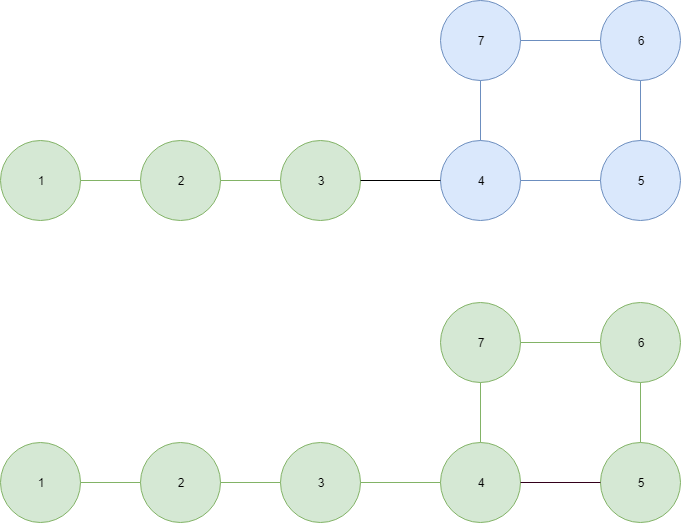}
				\caption{A path and a cycle are joined to obtain a new path.}
				\label{pathCycleJoin}       
			\end{figure*}
			
			\item The edge is between a cycle $C_i$ and a cycle $C_j$. The edge can be used to join the cycles if and only if $C_i$ is not the same cycle as $C_j$. In order to obtain a path by joining $C_i$ and $C_j$, the current edge is marked as active and furthermore one edge from $C_i$ and one edge from $C_j$ are marked as inactive. Again (as in case 2), there are two possible edges that can be marked as inactive for a given cycle and an arbitrary one of these two is chosen for each cycle. The set $S_1$ is updated by adding the path $P'$ that results from joining the cycles. The set $S_2$ is updated by removing $C_i$ and $C_j$.
		\end{enumerate}
		
		After having gone through all edges of $G$ and joining the structures at the endpoints of the edges in case this is possible, we have obtained a new set of paths $S_1'=\{P_1,P_2,\ldots,P_{k_1'}\}$ and a new set of cycles $S_2'=\{C_1,C_2,\ldots,C_{k_2' }\}$ such that $k_1'+k_2' \leq k$. Finally, for all cycles in $S_2$ an arbitrary edge is chosen to mark as inactive (to transform the cycle back into a path). The resulting set of active edges forms a (new) path partition of $G$.
		
		\item This step is identical to step 3 from section \ref{initialSpanningTree}. The path partition that is obtained from step 2 is transformed into another path partition by applying rotation moves to all of the paths of the path partition for which this is possible.
		
		\item This step is identical to step 4 from section \ref{initialSpanningTree}. The path partition from step 3 is transformed into a spanning tree $T'$ of $G$ by marking a particular set of edges as active.
	\end{enumerate}
	
	To analyse the time complexity of perturbing a spanning tree, we will analyse the time complexity of the different steps separately. Recall from lemma \ref{treeLemma} that step 1 can be implemented in $O(|V|)$. Step 2 can be implemented in $O(|V|+|E|)$, because the state of an edge will only be changed at most once (from active to inactive or vice-versa). As we have seen in section \ref{initialSpanningTree}, step 3 can be implemented in $O(|V|+|E|)$ and step 4 can be implemented in $O(|E|*\alpha(|V|))$. Hence, the total time complexity of perturbing a spanning tree is $O(|V|+|E|*\alpha(|V|))$. 
	
	\subsection{Complete algorithm}
	\label{completeAlgorithm}
	
	Now that the building blocks have been explained, we are ready to discuss the complete algorithm. The algorithm has three parameters: \textit{preferredRatio}, \textit{maxAmountInitialSpanningTrees} and \textit{maxAmountBadPerturbations}. The first parameter (\textit{preferredRatio}) has already been discussed in step 4 of section \ref{initialSpanningTree}, whereas \textit{maxAmountInitialSpanningTrees} indicates how many initial spanning trees have to be generated. The third parameter (\textit{maxAmountBadPerturbations}) requires a more detailled discussion of a crucial property of the perturbation operator.
	
	For every initial spanning tree $T$ of $G$, a number of perturbations will be performed. Let \textit{perturbed}$(T,G)$ denote the spanning tree that is obtained by perturbing the spanning tree $T$ (by making certain edges in $G$ active and others inactive). The perturbation that was discussed before has the property that $PPN($\textit{perturbed}$(T,G)) \leq PPN(T)$, so perturbing a spanning tree always yields another spanning tree that is at least as good. This property holds, because for every step of the perturbation, the path partition number of the graph that consists of the set of active edges does not increase. This property is the motivation for introducing the concept of a bad perturbation of a spanning tree: a perturbation such that $PPN($\textit{perturbed}$(T,G))=PPN(T)$. The parameter \textit{maxAmountBadPerturbations} indicates how many bad perturbations are applied to a spanning tree, before we move on to the next spanning tree. 
	
	For every spanning tree $T'$, that is obtained after performing perturbations on an initial spanning tree $T$, a minimum path partition is computed. This path partition is used to estimate $HCN(G)$ (see lemma \ref{HCNPPNLemma}). In case $PPN(T' )>1$, the algorithm estimates $HCN(G)$ to be equal to $PPN(T')$. In the special case where the path partition consists of a single path (i.e. $PPN(T' )=1$), it is verified whether there is an edge in $G$ between the endpoints of this path. The algorithm estimates $HCN(G)$ as $1$ in case such an edge does not exist and $0$ otherwise.
	
	The pseudocode of the complete algorithm is shown in algorithm 3. Here, \textit{estimateHCN} is a function that accepts as input a graph $G$ for which HCP has to be solved and produces as output an estimate for $HCN(G)$. The function \textit{obtainInitialSpanningTree} has been discussed in section \ref{initialSpanningTree}. The function \textit{computeMinimumPathPartitionOfTree} accepts as input a tree and produces as output a minimum path partition of that tree. The function \textit{perturbed} accepts as input a spanning tree of $G$ as well as the graph $G$ itself and produces as output a (new) spanning tree of $G$ (see section \ref{perturbing}). Finally, the function \textit{estimateHCNPathPartition} accepts as input a path partition of $G$ as well as the graph $G$ itself and produces as output an estimate of $HCN(G)$.
	
	\begin{algorithm}[h!]
		\FnC{\FRecursC{G}}{
			\KwData{\\G: an undirected graph}
			\KwResult{\\An estimate for $HCN(G)$}
			\tcc{Start of code}
			{$estimate \gets \infty$}\;
			\For{\forcond}{
				{$T \gets obtainInitialSpanningTree(G)$}\;
				{$amountBadPerturbations \gets 0$}\;
				{$pathPartition \gets computeMinimumPathPartitionOfTree(T)$}\;
				
				\While{$amountBadPerturbations \leq maxAmountBadPerturbations$}{
					{$T \gets perturbed(T,G)$}\;
					{$newPathPartition \gets computeMinimumPathPartitionOfTree(T)$}\;
					\uIf{$pathPartition$ and $newPathPartition$ have the same amount of paths}{
						{$amountBadPerturbations \gets amountBadPerturbations+1$}\;
					}
					{$pathPartition \gets newPathPartition$}\;
					{$newEstimate \gets estimateHCNPathPartition(pathPartition,G)$}\;
					\uIf{$newEstimate  < estimate$}{
						{$estimate \gets newEstimate$}\;
					}
					\If(\tcc*[h]{Guaranteed global optimum}){$estimate=0$}{
						{\Return 0}\;
					}
				}
				
			}
			{\Return $estimate$}\;
		}
		\caption{Estimate the HCN of a graph}
	\end{algorithm}
	
	To analyse the worst-case time complexity of the complete algorithm, first note that the path partition number of any graph $G=(V,E)$ is at most equal to $|V|$, because of the existence of a trivial path partition in which every path consists of a single node. In every iteration of the while loop either the size or the path partition decreases by at least one or the size of the path partition remains the same, in which case the amount of bad perturbations is incremented. If we combine this insight with the time complexities from the previous subsections, we obtain for the complete algorithm a worst-case time complexity of $O($\textit{maxAmountInitialSpanningTrees}$)*(LKHTime(V,E)+(|V|+$\textit{maxAmountBadPerturbations}$)*(|V|+|E|*\alpha(|V|)))$. If the parameters are regarded as constants, the time complexity becomes $LKHTime(V,E)+O(|V|*|E|*\alpha(|V|))$.
	
	\section{Benchmark set generation}
	\label{benchmark}
	
	To the best of our knowledge, up until now there was no standard benchmark set of problem instances already available for HCP on undirected graphs. In order to be able to test the quality of the proposed algorithm, we have constructed a benchmark set that consists of a variety of problem instances. A problem instance for HCP on undirected graphs is simply an undirected graph for which HCP has to be solved.
	
	We have constructed a benchmark set of problem instances that vary with respect to size, density, degree distribution and several other features. It is composed both of graphs that arise from practice as well as graphs that were generated according to theoretical models. For the former case, a number of graphs were selected from the instances of the tenth DIMACS challenge about graph partitioning and graph clustering \cite{Bader:2013}. Some of these graphs are directed graphs. These graphs are transformed into undirected graphs (by ignoring the direction of the edges) to obtain the problem instances. The other problem instances were generated by using several graph generators. Some of these generators are part of the library of the Stanford Network Analysis Platform (SNAP) \cite{Leskovec:2016}. This library contains many graph generators that have an underlying theoretical model. A brief overview of the six generators that were used to devise the benchmark set follows.
	
	\begin{enumerate}[Gener{a}tor 1:]
		\item This generator follows the Erd\H{o}s-R\'enyi model \cite{Erdos:1959}. This graph generator is a stochastic generator and has two parameters $n$ and $p$. The generated graph consists of n nodes and for every pair of nodes there is an edge between them with probability $p$. Hence, each (simple, undirected) graph that consists of n nodes and m edges is equally likely to be generated by this model (with probability $p^{m}*(1-p)^{\frac{n*(n-1)}{2}-m}$). 
		
		The problem instances that were generated by this generator contain a varying amount of nodes (powers of 2 between $2^8$ and $2^{13}$). For a fixed amount of nodes $n$, several problem instances were generated with varying probabilities (such that the expected average degree of a node varies between $2^0, 2^1, 2^2, \ldots, n$).
		
		For graphs generated with this model, many theoretical results are known. We highlight two relevant theoretical results from \cite{Bollobas:1998}:
		\begin{theorem}
			Let $0<r<1$ be a real number and let $G(n,p)=(V,E)$ be a graph obtained by the Erd\H{o}s-R\'enyi random model with parameters $n$ and $p$. If $n$ tends to infinity and $p$ is chosen such that $|E|=o(|V|^r)$, the probability that all components of the graph are trees, tends to 1.
		\end{theorem}
		
		Because of the previous theorem, for sufficiently small values of $p$, the probability that the algorithm proposed in this paper computes the global optimum tends to 1 if n tends to infinity. This is the case, because HCP is solved exactly for trees.
		
		The following theorem indicates for which graphs the Hamiltonian completion number is likely to be equal to 0:
		\begin{theorem}
			Let $f(n)$ be a function that tends to infinity if $n$ tends to infinity and let $G(n,p)=(V,E)$ be a graph obtained by the Erd\H{o}s-R\'enyi random model with parameters $n$ and $p$. Set $p_l=\frac{log(n)+log(log(n))-f(n)}{n}$ and $p_u=\frac{log(n)+log(log(n))+f(n)}{n}$. If $n$ tends to infinity, the probability that $G(n,p_l)$ contains a Hamiltonian cycle tends to 0 and the probability that $G(n,p_u)$ contains a Hamiltonian cycle tends to 1.
		\end{theorem}
		
		Hence, for sufficiently large values of $p$ the generated graphs are verly likely to have a Hamiltonian completion number equal to 0.

		\item This generator generates circular graphs and has two parameters $n$ and $k$. The generated graph consists of $n$ nodes, labelled $0,1,\ldots,n-1$. For every node with label $i$, there is an edge to the $k$ subsequent nodes $(i+1) \bmod n, (i+2) \bmod n,\ldots, (i+k) \bmod n$. For all of these graphs, the Hamiltonian completion number is equal to $0$.
		
		The parameters that were used to generate problem instances by this generator are varying: $n$ is between $500$ and $30,000$ and $k$ is between $3$ and $10$.
		
		\item This generator generates two-dimensional grid graphs and has two parameters $n$ and $m$. The generated graph consists of $n*m$ nodes, which can be thought of as $n$ rows of $m$ nodes which are laid out in a two-dimensional grid. There is an edge between a node in row $r_1$ and column $c_1$ and a node in row $r_2$ and column $c_2$ if and only if $|r_1-r_2 |+|c_1-c_2 |=1$. For these graphs, the Hamiltonian completion number is equal to $0$ if either $n$ or $m$ are even and otherwise equal to $1$.
		
		The problem instances that were generated by this generator contain between $300$ and $10,000$ nodes.
		\item This generator generates preferential attachment graphs, following the Barab\'asi-Albert model \cite{Barabasi:1999}. In this model, the degree distribution of the nodes follows a power-law: the probability that a node is connected to $k$ other nodes is proportional with $ k^{-\gamma}$, where $\gamma$ is a parameter of the generator (to be set by the user).
		
		The graphs that were generated according to this model contain between $300$ and $3000$ nodes and the nodes have an average degree varying between $6$ and $16$.
		\item This generator starts from a star graph consisting of $n+1$ nodes and adds $n$ random edges to obtain a problem instance. A star graph consisting of $n+1$ nodes is the complete bipartite graph $K_{1,n}$ (a tree with $1$ internal node and $n$ leaves). 
		
		The problem instances that were generated by this generator contain between 1000 and 4000 nodes.
		
		\item The final generator generates structured trees and has two parameters $l$ and $c$. The generated graphs are trees that consist of $l$ levels of nodes and every node (except for the leave nodes) has $c$ child nodes. Note that for these graphs, the algorithm proposed in the paper is guaranteed to compute the global optimum, because our algorithm solves HCP exactly for trees.
		
		In order to not get results that are biased positively towards our own algorithm, only 2 problem instances were generated by this generator (one instance was generated with $l=7$ and $c=3$ and the other one with $l=10$ and $c=2$).
	\end{enumerate}
	
	\section{Results and experiments}
	\label{results}
	
	In total, the benchmark set consists of 113 problem instances. This benchmark set is split into two parts: a tuning set (consisting of 50 randomly chosen problem instances) and a test set (consisting of the remaining 63 problem instances). Table \ref{tableProblemDistribution} shows the distribution of the problem instances over each of the two sets. The tuning set will be used to find appropriate parameter settings for the algorithm proposed in this paper. Furthermore, we will demonstrate the effect of varying these parameters and how this influences the performance of the algorithm. The test set will be used to show the added value of the algorithm components that were discussed in this paper. Furthermore, we compare our algorithm with the performance of state-of-the-art algorithms and we attribute special attention to when our algorithm performs well and when it does not. All experiments were performed on a Intel i7-8550U CPU with a clock rate of 1.8 GHz. The test data is available online at \textit{https://set.kuleuven.be/codes/hamiltoniancompletionproblem} and at \textit{https://github.com/JorikJooken/ProblemInstancesHamiltonianCompletionProblem}.
	
	\begin{table}[h!] \footnotesize\centering 
		\begin{threeparttable}
			\caption{Amount of problems in tuning set and test set.} 
			\label{tableProblemDistribution} 
			\begin{tabular}{ D{.}{.}{-2} D{.}{.}{-2} D{.}{.}{-2} D{.}{.}{-2} D{.}{.}{-2}} \\
				\hline\noalign{\smallskip} 
				\multicolumn{1}{l}{} & \multicolumn{1}{c}{Tuning set} & \multicolumn{1}{c}{Test set}\\ 
				\noalign{\smallskip}
				\hline
				\noalign{\smallskip} 
				\multicolumn{1}{l}{DIMACS} & 8 & 7 \\ 
				\multicolumn{1}{l}{Erd\H{o}s-R\'enyi} & 29 & 39  \\ 
				\multicolumn{1}{l}{circular} & 3 & 6 \\ 
				\multicolumn{1}{l}{grid} & 5 & 4\\ 
				\multicolumn{1}{l}{preferential attachment} & 3 & 4 \\ 
				\multicolumn{1}{l}{star} & 2 & 1  \\ 
				\multicolumn{1}{l}{structured tree} & 0 & 2 \\ 
				\noalign{\smallskip}
				\hline
				\noalign{\smallskip}
			\end{tabular}
		\end{threeparttable}
	\end{table}
	
	\subsection{Tuning}
	\label{tuning}
	
	As discussed before, the algorithm has three parameters: \textit{preferredRatio}, \textit{maxAmountInitialSpanningTrees} and \textit{maxAmountBadPerturbations}. For the parameter \textit{preferredRatio}, it is not immediately clear without any experiments which setting would lead to the best performance. For the other two parameters, however, it is clear that higher values correspond to a better performing algorithm, because they indicate the amount of local search that has to be done. Hence, for these two parameters we will try to find settings that yield a good trade-off between execution time and performance. Note that the LKH solver that is used in step 1 of the algorithm also has parameters itself. These parameters will not be subject to tuning, but instead they are chosen in such a way that only a fraction of the full power of the LKH solver is being used. The precise parameters that were supplied to this solver can be found in table \ref{tableLKHParameters}.
	
	\begin{table}[h!] \footnotesize\centering
		\begin{threeparttable}
			\caption{Parameter settings for LKH for step 1 of constructing an initial spanning tree} 
			\label{tableLKHParameters} 
			\begin{tabular}{D{.}{.}{-2} D{.}{.}{-2} D{.}{.}{-2} D{.}{.}{-2} D{.}{.}{-2} } \\
				\hline\noalign{\smallskip} 
				\multicolumn{1}{l}{Parameter} & \multicolumn{1}{c}{Value} \\ 
				\noalign{\smallskip}
				\hline
				\noalign{\smallskip} 
				\multicolumn{1}{l}{$MOVE\_TYPE$} & 5 \\ 
				\multicolumn{1}{l}{$PATCHING\_C$} & 3  \\ 
				\multicolumn{1}{l}{$PATCHING\_A$} & 2 \\ 
				\multicolumn{1}{l}{\textit{RUNS}} & 1 \\ 
				\multicolumn{1}{l}{$CANDIDATE\_SET\_TYPE$} & \textit{POPMUSIC} \\ 
				\multicolumn{1}{l}{$MAX\_TRIALS$} & 2 \\ 
				\noalign{\smallskip}
				\hline
				\noalign{\smallskip}
			\end{tabular}
		\end{threeparttable}
	\end{table}
	
	In the following sections, the multi-start local search algorithm from this paper will be used with various parameter settings. For the sake of notational convenience, we will abbreviate the multi-start local search algorithm with a specific parameter setting as $MSLS\_{a}\_{b}\_{c}$, where $a$, $b$ and $c$ correspond to the chosen values for \textit{preferredRatio}, \textit{maxAmountInitialSpanningTrees} and \textit{maxAmountBadPerturbations} respectively. For every problem instance, the algorithm is interrupted after 1000 seconds if the algorithm has not finished within this time. If the algorithm finishes in less than 1000 seconds, no further search is performed to fill the remaining time. In order to obtain fair comparisons, in all of the experiments the reported averages will be computed only over those problem instances for which all algorithms (in the scope of the specific experiment) were able to produce an answer within 1000 seconds. Hence, the reported averages can vary significantly between different experiments (depending on which problem instances the averages are computed over). This makes the reported averages only comparable within one specific experiment. We will report separately about the problem instances for which an algorithm did not produce an answer within 1000 seconds. All the execution times reported in this paper measure the time that passes between the start of the algorithm and the end of the algorithm, unless stated otherwise. Note that this is different from the time that passes between the start of the algorithm and the first moment at which the reported result is found.
	
	In the first experiment, we have used several values (1, 5, 25 and 125) for the parameter \textit{preferredRatio}. The summary of these results can be found in table \ref{tableVaryingPreferredRatio}. The first row of the table indicates on how many problem instances the algorithm did not finish within 1000 seconds, whereas the second row of the table shows the average amount of time needed to solve a problem instance. The third row displays the average amount of edges that the algorithm has added in order to obtain a Hamiltonian graph (the smaller the better). From this table, we can see that varying this parameter has little effect (both regarding the amount of interruptions, the average execution time and the average amount of edges). The parameter setting of \textit{preferredRatio} $=25$ has the best result for both average execution time and the average amount of edges. As a result, we will use this value in the future experiments.
	
	\begin{table}[h!] \footnotesize\centering 
		\begin{threeparttable}
			\caption{Results for various values of \textit{preferredRatio}} 
			\label{tableVaryingPreferredRatio} 
			\begin{tabular}{ D{.}{.}{-2} D{.}{.}{-2} D{.}{.}{-2} D{.}{.}{-2} D{.}{.}{-2}} \\
				\hline\noalign{\smallskip} 
				\multicolumn{1}{l}{} & \multicolumn{1}{c}{$MSLS\_1\_10\_3000$} & \multicolumn{1}{c}{$MSLS\_5\_10\_3000$} & \multicolumn{1}{c}{$MSLS\_25\_10\_3000$} & \multicolumn{1}{c}{$MSLS\_125\_10\_3000$}\\ 
				\noalign{\smallskip}
				\hline
				\noalign{\smallskip} 
				\multicolumn{1}{l}{amount interrupted} & 1 & 1 & 1 & 0 \\ 
				\multicolumn{1}{l}{average time (in seconds)} & 83.19 & 79.31 & 69.56 & 74.44  \\ 
				\multicolumn{1}{l}{average amount edges added} & 234.53 & 234.51 & 234.51 & 234.55 \\ 
				\noalign{\smallskip}
				\hline
				\noalign{\smallskip}
			\end{tabular}
		\end{threeparttable}
	\end{table}
	
	The second experiment that we have performed, concerns the effect of changing the parameter \textit{maxAmountInitialSpanningTrees}. The results of this experiment are summarised in table \ref{tableVaryingMaxAmountInitialSpanningTrees}. From this table it is clear that increasing \textit{maxAmountInitialSpanningTrees} does not improve the average amount of edges added a lot. The average computing time does, however, increase quite fast. Note that the computing time is multiplied by less than a factor of 10 if \textit{maxAmountInitialSpanningTrees} is multiplied by 10. This happens, because the algorithm can stop considering new spanning trees as soon as a spanning tree with a HCN of 0 is found. There are 8 problem instances for which the algorithm was not able to finish within 1000 seconds if \textit{maxAmountInitialSpanningTrees} is equal to 100, but only 1 such problem instance if this parameter is equal to 10. This leads us to use a value of 10 for \textit{maxAmountInitialSpanningTrees}.
	
	\begin{table}[h!] \footnotesize\centering 
		\begin{threeparttable}
			\caption{Results for various values of \textit{maxAmountInitialSpanningTrees}} 
			\label{tableVaryingMaxAmountInitialSpanningTrees} 
			\begin{tabular}{ D{.}{.}{-2} D{.}{.}{-2} D{.}{.}{-2} D{.}{.}{-2} D{.}{.}{-2}} \\
				\hline\noalign{\smallskip} 
				\multicolumn{1}{l}{} & \multicolumn{1}{c}{$MSLS\_25\_1\_3000$} & \multicolumn{1}{c}{$MSLS\_25\_10\_3000$} & \multicolumn{1}{c}{$MSLS\_25\_100\_3000$}\\ 
				\noalign{\smallskip}
				\hline
				\noalign{\smallskip} 
				\multicolumn{1}{l}{amount interrupted} & 0 & 1 & 8 \\ 
				\multicolumn{1}{l}{average time (in seconds)} & 11.26 & 26.33 & 121.53  \\ 
				\multicolumn{1}{l}{average amount edges added} & 39.10 & 39.00 & 38.98 \\ 
				\noalign{\smallskip}
				\hline
				\noalign{\smallskip}
			\end{tabular}
		\end{threeparttable}
	\end{table}
	
	Finally, we performed an experiment where we have used several values (100, 300, 1000 and 3000) for the parameter \textit{maxAmountBadPerturbations}. The results of this experiment are summarised in table \ref{tableVaryingMaxAmountBadPerturbations}. Varying this parameter has a larger impact on the result (average amount of edges added) than the previous two parameters. By increasing this parameter, the amount of edges to add gets better (lower) and the execution time gets worse (higher). Note, however, that again (as was the case for \textit{maxAmountInitialSpanningTrees}) the ratio between the execution time and \textit{maxAmountBadPerturbations} decreases if \textit{maxAmountBadPerturbations} increases. If we choose a value of 3000 for \textit{maxAmountBadPerturbations}, there are quite some problem instances for which the execution time is close to 1000 seconds. Hence, we have chosen to not increase this parameter any further and we will use the value of 3000.
	
	\begin{table}[h!] \footnotesize\centering 
		\begin{threeparttable}
			\caption{Results for various values of \textit{maxAmountBadPerturbations}} 
			\label{tableVaryingMaxAmountBadPerturbations} 
			\begin{tabular}{ D{.}{.}{-2} D{.}{.}{-2} D{.}{.}{-2} D{.}{.}{-2} D{.}{.}{-2}} \\
				\hline\noalign{\smallskip} 
				\multicolumn{1}{l}{} & \multicolumn{1}{c}{$MSLS\_25\_10\_100$} & \multicolumn{1}{c}{$MSLS\_25\_10\_300$} & \multicolumn{1}{c}{$MSLS\_25\_10\_1000$} & \multicolumn{1}{c}{$MSLS\_25\_10\_3000$}\\ 
				\noalign{\smallskip}
				\hline
				\noalign{\smallskip} 
				\multicolumn{1}{l}{amount interrupted} & 0 & 1 & 1 & 1 \\ 
				\multicolumn{1}{l}{average time (in seconds)} & 11.52 & 20.89 & 44.48 & 69.50  \\ 
				\multicolumn{1}{l}{average amount edges added} & 237.55 & 236.08 & 235.08 & 234.51 \\ 
				\noalign{\smallskip}
				\hline
				\noalign{\smallskip}
			\end{tabular}
		\end{threeparttable}
	\end{table}
	
	These experiments lead us to use the following parameter configuration for testing purposes: \textit{preferredRatio} $=25$, \textit{maxAmounInitalSpanningTrees} $=10$ and \textit{maxAmountBadPerturbations} $=3000$.
	
	\subsection{Contribution of algorithm components}
	\label{contribution}
	
	The algorithm proposed in this paper makes use of the TSP solver LKH (in step 1 of the process of obtaining an initial spanning tree). In this subsection, we show that the other algorithm components contribute significantly to the final result obtained by our algorithm. We make a comparison between the results obtained by on the one hand $MSLS\_25\_10\_3000$ and on the other hand an algorithm that only performs step 1 of the process of obtaining an initial spanning tree (i.e. $MSLS\_25\_10\_3000$ without all other steps). We will call the latter algorithm $MULTI\_LKH$. The same parameter settings were used for LKH in both algorithms. These parameter settings can be found in table \ref{tableLKHParameters}. 
	
	The results of running these two algorithms on the problem instances of the test set can be found in table \ref{tableContributionComparison}. From this table, we can see that about $20 \%$ of the total computing time of $MSLS\_25\_10\_3000$ was spent on step 1 for obtaining initial spanning trees. By also performing the other steps, the amount of added edges could on average be reduced by approximately 7. Hence, the contribution of the other steps is significant. This is also confirmed by doing a paired samples t-test on the pairs obtained by pairing the results of both algorithms. The null hypothesis in this test states that the mean difference (the result of $MSLS\_25\_10\_3000$ minus the result of $MULTI\_LKH$) is equal to 0 and the alternative hypothesis states that the mean difference is smaller than 0. A p-value of 0.004 was obtained, such that at a significance level of $1 \%$, we reject the null hypothesis in favor of the alternative hypothesis.
	
	\begin{table}[h!] \footnotesize\centering 
		\begin{threeparttable}
			\caption{Results on test set for $MSLS\_25\_10\_3000$ and for $MULTI\_LKH$} 
			\label{tableContributionComparison} 
			\begin{tabular}{ D{.}{.}{-2} D{.}{.}{-2} D{.}{.}{-2} D{.}{.}{-2} D{.}{.}{-2}} \\
				\hline\noalign{\smallskip} 
				\multicolumn{1}{l}{} & \multicolumn{1}{c}{$MSLS\_25\_10\_3000$} & \multicolumn{1}{c}{$MULTI\_LKH$} \\ 
				\noalign{\smallskip}
				\hline
				\noalign{\smallskip} 
				\multicolumn{1}{l}{amount interrupted} & 0 & 0 \\ 
				\multicolumn{1}{l}{average time (in seconds)} & 50.58 & 9.96 \\ 
				\multicolumn{1}{l}{average amount edges added} & 158.44 & 165.40 \\ 
				\noalign{\smallskip}
				\hline
				\noalign{\smallskip}
			\end{tabular}
		\end{threeparttable}
	\end{table}
	
	\subsection{Comparison with existing solvers}
	\label{testing}
	
	We will compare the results of our algorithm with the results obtained by the TSP solvers LKH and Concorde by regarding HCP as a specific TSP instance (see subsection \ref{reductionTSP}). Concorde is an exact solver and hence for this solver the most interesting aspect is its computation time. LKH on the other hand is a heuristic solver. For LKH, we have used the recommended parameter settings \cite{Helsgaun:2000} in this experiment. Note that these parameter settings are much more time consuming than those used by LKH in step 1 of our own algorithm. The parameter settings used for LKH in this experiment can be found in table \ref{tableLKHParameters2}.
	
	\begin{table}[h!] \footnotesize\centering
		\begin{threeparttable}
			\caption{Recommended parameter settings for LKH} 
			\label{tableLKHParameters2} 
			\begin{tabular}{D{.}{.}{-2} D{.}{.}{-2} D{.}{.}{-2} D{.}{.}{-2} D{.}{.}{-2} } \\
				\hline\noalign{\smallskip} 
				\multicolumn{1}{l}{Parameter} & \multicolumn{1}{c}{Value} \\ 
				\noalign{\smallskip}
				\hline
				\noalign{\smallskip} 
				\multicolumn{1}{l}{$MOVE\_TYPE$} & 5 \\ 
				\multicolumn{1}{l}{$PATCHING\_C$} & 3  \\ 
				\multicolumn{1}{l}{$PATCHING\_A$} & 2 \\ 
				\multicolumn{1}{l}{$RUNS$} & 10 \\ 
				\multicolumn{1}{l}{$CANDIDATE\_SET\_TYPE$} & \textit{POPMUSIC} \\ 
				\multicolumn{1}{l}{$MAX\_TRIALS$} & |V| \\ 
				\noalign{\smallskip}
				\hline
				\noalign{\smallskip}
			\end{tabular}
		\end{threeparttable}
	\end{table}
	
	The results of running these three algorithms on the 63 problem instances of the test set are summarised in table \ref{tableComparison}. All three algorithms were given a maximum computation time of 1000 seconds per test instance. The multi-start local search algorithm proposed in this paper needed on average around $72.7 \%$ and $36.2 \%$ of the time needed by Concorde and LKH respectively. The average amount of edges added produced by the three algorithms are very close to each other (Concorde produced the best results, followed by $MSLS\_25\_10\_3000$, followed by LKH). Recall that Concorde is an exact solver, so the result achieved by Concorde is the global optimum. From this comparison, we see that the other two algorithms are also often able to compute a solution that is (close to) the global optimum. 
	
	\begin{table}[h!] \footnotesize\centering 
		\begin{threeparttable}
			\caption{Results for Concorde, $MSLS\_25\_10\_3000$ and LKH} 
			\label{tableComparison} 
			\begin{tabular}{ D{.}{.}{-2} D{.}{.}{-2} D{.}{.}{-2} D{.}{.}{-2} D{.}{.}{-2}} \\
				\hline\noalign{\smallskip} 
				\multicolumn{1}{l}{} & \multicolumn{1}{c}{Concorde} & \multicolumn{1}{c}{$MSLS\_25\_10\_3000$} & \multicolumn{1}{c}{LKH}\\ 
				\noalign{\smallskip}
				\hline
				\noalign{\smallskip} 
				\multicolumn{1}{l}{amount interrupted} & 4 & 0 & 6 \\ 
				\multicolumn{1}{l}{average time (in seconds)} & 42.38 & 30.82 & 85.24  \\ 
				\multicolumn{1}{l}{average amount edges added} & 184.28 & 184.56 & 184.70 \\ 
				\noalign{\smallskip}
				\hline
				\noalign{\smallskip}
			\end{tabular}
		\end{threeparttable}
	\end{table}
	
	There were 51 problem instances (out of 63) for which all three algorithms were able to produce the global optimum. The results on the other problem instances are shown in table \ref{tableProblemInstances}. The values in the table are the amount of edges added by each algorithm, whereas the corresponding computation times are provided between brackets. $MSLS\_25\_10\_3000$ was able to compute an answer within 1000 seconds for all problem instances, while Concorde and LKH did not finish within this time limit for 4 and 6 problem instances respectively (indicated by the dashes in the table). The results obtained by $MSLS\_25\_10\_3000$ were at least as good as those obtained by LKH for all problem instances except for one problem instance: a grid graph consisting of 2 rows and 5000 columns ($grid\_graph\_2\_5000$ in table \ref{tableProblemInstances}).
	The problem instances \textit{preferential}$\_$\textit{attachment}$\_1500\_4$ and \textit{preferential}$\_$\textit{attachment}$\_2000\_4$ deserve further attention. These problem instances are preferential attachment graphs consisting of 1500 and 2000 nodes respectively. The average degree of a node is 8 for both problem instances. These two problem instances are the only problem instances for which the results obtained by $MSLS\_25\_10\_3000$ and LKH differ significantly from the global optimum. The global optimum for \textit{preferential}$\_$\textit{attachment}$\_1500\_4$ is 1 and $MSLS\_25\_10\_3000$ and LKH produced an answer of 9 and 12 respectively. Similarly the global optimum is 0 for \textit{preferential}$\_$\textit{attachment}$\_2000\_4$, while $MSLS\_25\_10\_3000$ and LKH produced an answer of 7 and 10 respectively. We suspect that the reason why these particular instances are hard to solve for $MSLS\_25\_10\_3000$ and LKH is related to the degree distribution of the nodes. If the nodes in a graph all have a very low degree, the graph can be thought of as being quite similar to a tree, because the average degree of a node in a tree is approximately equal to 2. Hence, one could expect that for these graphs HCP can be solved (nearly) optimal, because HCP can be solved exactly in linear time for trees. If, however, the nodes in a graph all have a very high degree, the graph can be expected to contain many Hamiltonian cycles, such that the Hamiltonian completion number of these graphs will most likely be close to 0. This implies that one could also expect that the HCP can be solved to (near-)optimality for these graphs. The degree distribution for the nodes in preferential attachment graphs does, however, not belong to either of these two categories. Preferential attachment graphs have a power-law degree distribution and hence most nodes have a small degree while only a few nodes have a very large degree in comparison with the other nodes.
	
	\begin{table}[h!] \footnotesize \centering 
		\begin{threeparttable}
			\caption{Results on problem instances for which at least one algorithm did not produce the global optimum. A dash indicates that the algorithm was not finished within 1000 seconds.}
			\label{tableProblemInstances} 
			\begin{tabular}{ D{.}{.}{-2} D{.}{.}{-2} D{.}{.}{-2} D{.}{.}{-2} D{.}{.}{-2}} \\
				\hline\noalign{\smallskip} 
				\multicolumn{1}{l}{Name problem instance} & \multicolumn{1}{c}{Concorde} & \multicolumn{1}{c}{$MSLS\_25\_10\_3000$} & \multicolumn{1}{c}{LKH}\\ 
				\noalign{\smallskip}
				\hline
				\noalign{\smallskip} 
				\multicolumn{1}{l}{$circle\_like\_10000\_3$} & $0 (44.35 s)$ & $0 (12.12 s)$ & -\\
				\multicolumn{1}{l}{$circle\_like\_15000\_3$} & $0 (75.35 s)$ & $0 (17.63 s)$ & -\\
				\multicolumn{1}{l}{$circle\_like\_30000\_3$} & $0 (433.76 s)$ & $0 (21.14 s)$ & -\\
				\multicolumn{1}{l}{$delaunay\_n12$} & - & $0 (140.34 s)$ & $2 (963.57 s)$\\
				\multicolumn{1}{l}{$er\_10\_2$} & $78 (6.31 s)$ & $78 (40.81 s)$ & $80 (26.9 s)$\\
				\multicolumn{1}{l}{$er\_13\_11$} & $0 (101.3 s)$ & $0 (347.38 s)$ & -\\
				\multicolumn{1}{l}{$er\_13\_3$} & $12 (78.02 s)$ & $12 (573.89 s)$ & -\\
				\multicolumn{1}{l}{$grid\_graph\_2\_5000$} & - & $3 (188.04 s)$ & $0 (33.85 s)$\\
				\multicolumn{1}{l}{$grid\_graph\_50\_50$} & - & $0 (22.48 s)$ & $4 (154.79 s)$\\
				\multicolumn{1}{l}{$grid\_graph\_80\_80$} & - & $1 (199.02 s)$ & -\\
				\multicolumn{1}{l}{\textit{preferential}$\_$\textit{attachment}$\_1500\_4$} & $1 (17.67 s)$ & $9 (75.18 s)$ & $12 (279.45 s)$\\
				\multicolumn{1}{l}{\textit{preferential}$\_$\textit{attachment}$\_2000\_4$} & $0 (16.5 s)$ & $7 (107.53 s)$ & $10 (367.8 s)$\\
				\noalign{\smallskip}
				\hline
				\noalign{\smallskip}
			\end{tabular}
		\end{threeparttable}
	\end{table}
	Finally, we feel the need to point out that it is possible to change the experimental setup such that there exist parameter settings for the LKH solver that are able to obtain better results than the recommended settings. These improved parameter settings can be found in table \ref{tableLKHParameters3}. Recall that the execution times measured in the rest of the paper measure the time that passes between the start of the algorithm and the end of the algorithm. By doing this, we give the algorithm the opportunity to be able to compute the best possible answer within the specified amount of time of 1000 seconds and hence the algorithm is not interrupted if no improvement has been found after a long time. If we instead measure the time that passes between the start of the algorithm and the first moment upon which the reported answer is found, we obtain a different view. The results in this new experimental setting on the same problem instances as in table \ref{tableProblemInstances} can be found in table \ref{tableProblemInstances2}. These results were kindly contributed to us by the creator of LKH, Keld Helsgaun. This experiment was performed on an iMac with a clock rate of 3.4 GHz.
	
	\begin{table}[h!] \footnotesize\centering
		\begin{threeparttable}
			\caption{Non-default parameter settings for LKH} 
			\label{tableLKHParameters3} 
			\begin{tabular}{D{.}{.}{-2} D{.}{.}{-2} D{.}{.}{-2} D{.}{.}{-2} D{.}{.}{-2} } \\
				\hline\noalign{\smallskip} 
				\multicolumn{1}{l}{Parameter} & \multicolumn{1}{c}{Value} \\ 
				\noalign{\smallskip}
				\hline
				\noalign{\smallskip} 
				\multicolumn{1}{l}{$MAX\_CANDIDATES$} & 10 \\ 
				\multicolumn{1}{l}{$SUBGRADIENT$} & NO  \\ 
				\multicolumn{1}{l}{$RESTRICTED\_SEARCH$} & NO \\ 
				\noalign{\smallskip}
				\hline
				\noalign{\smallskip}
			\end{tabular}
		\end{threeparttable}
	\end{table}
	
	\begin{table}[h!] \footnotesize \centering 
		\begin{threeparttable}
			\caption{Results produced by LKH with non-default parameter settings. The execution times measure the time that passes between the start of the algorithm and the first moment upon which the reported result is found. This is different from the other times reported in this paper.}
			\label{tableProblemInstances2} 
			\begin{tabular}{ D{.}{.}{-2} D{.}{.}{-2} D{.}{.}{-2} D{.}{.}{-2} D{.}{.}{-2}} \\
				\hline\noalign{\smallskip} 
				\multicolumn{1}{l}{Name problem instance} & \multicolumn{1}{c}{LKH}\\ 
				\noalign{\smallskip}
				\hline
				\noalign{\smallskip} 
				\multicolumn{1}{l}{$circle\_like\_10000\_3$} & $0 (11.63 s)$ \\
				\multicolumn{1}{l}{$circle\_like\_15000\_3$} & $0 (28.74 s)$ \\
				\multicolumn{1}{l}{$circle\_like\_30000\_3$} & $0 (143.02 s)$\\
				\multicolumn{1}{l}{$delaunay\_n12$} & $0 (1.98 s)$ \\
				\multicolumn{1}{l}{$er\_10\_2$} & $78 (0.12 s)$\\
				\multicolumn{1}{l}{$er\_13\_11$} & $0 (7.90 s)$\\
				\multicolumn{1}{l}{$er\_13\_3$} & $12 (9.81 s)$\\
				\multicolumn{1}{l}{$grid\_graph\_2\_5000$} & $0 (12.17 s)$\\
				\multicolumn{1}{l}{$grid\_graph\_50\_50$} & $0 (3.56 s)$\\
				\multicolumn{1}{l}{$grid\_graph\_80\_80$} & $0 (86.70 s)$\\
				\multicolumn{1}{l}{\textit{preferential}$\_$\textit{attachment}$\_1500\_4$} & $1 (21.38 s)$\\
				\multicolumn{1}{l}{\textit{preferential}$\_$\textit{attachment}$\_2000\_4$} & $0 (1.08 s)$\\
				\noalign{\smallskip}
				\hline
				\noalign{\smallskip}
			\end{tabular}
		\end{threeparttable}
	\end{table}
	
	\section{Conclusions and further work}
	\label{conclusionsFurtherWork}
	
	In this paper, we have proposed a multi-start local search algorithm for HCP on undirected graphs. Furthermore, we have constructed a benchmark set of problem instances for this problem, which can hopefully help future researchers to be able to compare the performance of their algorithms against other algorithms. This benchmark set was used to demonstrate that the algorithm in the paper performs well in comparison with state-of-the-art TSP solvers, both in respect of time and quality of the produced answer.
	
	An interesting idea that was not further explored in this paper is related to decomposing a problem instance into smaller problem instances. For HCP, the articulation points and bridges of the graph allow it to be decomposed, as discussed in \cite{Rayward:1987}. It is, however, not immediately clear whether the overhead involved in decomposing the problem and reassembling the different parts of the solution is worthwhile.
	
	An other interesting future research avenue is inspired upon the work that has been done for TSP. The LKH TSP solver first converts the input graph to a new graph in which some edges, that are unlikely to be part of an optimal TSP tour, are removed. By doing this, the search space can be drastically reduced and the solver can focus its search on more promising parts of the search space. The most common measure that is used to estimate how likely it is that a certain edge is part of an optimal TSP tour is called the $\alpha$-nearness of an edge \cite{Helsgaun:2000}. Note that for HCP the $\alpha$-nearness of all edges are equal and hence this measure is not really useful for HCP. It would be interesting to see whether some similar measure could be used for HCP.
	
	\section*{Acknowledgements} We gratefully acknowledge the support provided by the ORDinL project (FWO-SBO S007318N, Data Driven Logistics, 1/1/2018 - 31/12/2021). Pieter Leyman is a Postdoctoral Fellow of the Research Foundation - Flanders (FWO) with contract number 12P9419N.
	
	We also gratefully acknowledge the contribution of the data from table \ref{tableLKHParameters3} and table \ref{tableProblemInstances2} by Keld Helsgaun.

This is a pre-print of an article published in Journal of Heuristics. The final authenticated version is available online at: https://doi.org/10.1007/s10732-020-09447-9.
	\small
	\bibliographystyle{abbrv}      
	\bibliography{references}   
	
\end{document}